\documentclass[aps,floatfix,showpacs,showkeys,amsmath]{revtex4-1}
\usepackage{color,graphicx}
\usepackage{dcolumn}
\usepackage{bm}
\usepackage{epsfig}
\usepackage{supertabular}
\usepackage[bookmarksopen]{hyperref}
\def\be {\begin{equation}}
\def\ee {\end{equation}}
\def\nn {\nonumber}
\def\bea {\begin{eqnarray}}
\def\eea {\end{eqnarray}}

\def  \g    {\gamma}

\def  \p    {\pi}

\def  \m    {\mu}

\def  \f    {\frac}
\def  \lt   {\left}
\def  \rt   {\right}

\def  \lt   {\left}
\def  \rt   {\right}

\def  \del  {\partial}

\def  \ep   {\epsilon}

\def  \bef  {\begin{figure}}
\def  \eef  {\end{figure}}
\def  \be   {\begin{equation}}
\def  \ee   {\end{equation}}
\def  \ba   {\begin{array}}
\def  \ea   {\end{array}}
\def  \bea  {\begin{eqnarray}}
\def  \eea  {\end{eqnarray}}
\def  \beq  {\begin{eqnarray}}
\def  \eeq  {\end{eqnarray}}
\def  \nn   {\nonumber}
\def  \bd   {\begin{displaymath}}
\def  \ed   {\end{displaymath}}
\def  \bse  {\begin{subequations}}
\def  \ese  {\end{subequations}}
\def  \bwt  {\begin{widetext}}
\def  \ewt  {\end{widetext}}

\def  \ba   {{\bf{a_1}}}

\begin{document} 
 \title{Pionic dispersion relations in the presence of a weak magnetic field}
\bigskip
\bigskip

\author{Souvik Priyam Adhya}
 \affiliation{High Energy Nuclear and Particle Physics Division, Saha Institute
of Nuclear Physics,
1/AF Bidhannagar, Kolkata-700 064, INDIA}
 \email{souvikpriyam.adhya@saha.ac.in}

 \author{Mahatsab Mandal}
 \affiliation{High Energy Nuclear and Particle Physics Division, Saha Institute
of Nuclear Physics,
1/AF Bidhannagar Kolkata-700 064, INDIA}
\email{mahatsab@gmail.com}

\author{Subhrajyoti Biswas}
\affiliation{
Department of Physics, Rishi Bankim Chandra College,
Naihati, West Bengal - 743165, India}
\email{anjansubhra@gmail.com}

 \author{Pradip K. Roy}
 \email{pradipk.roy@saha.ac.in}
\affiliation{High Energy Nuclear and Particle Physics Division, Saha Institute
of Nuclear Physics,
1/AF Bidhannagar, Kolkata-700 064 INDIA}

\medskip

\begin{abstract}
In this work, dispersion relations of 
$\pi^0$ and $\pi^{\pm}$ have been studied
in vacuum in the limit of weak external magnetic field using
a phenomenological pion-nucleon $(\pi N)$ Lagrangian. 
For our purpose, we have calculated the results up to one loop order 
in self energy diagrams with the pseudoscalar $(PS)$ and 
 pseudovector $(PV)$ pion-nucleon interactions. 
By assuming weak external magnetic field it is seen that the effective mass
of pion gets explicit 
magnetic field dependence and it is modified significantly for the case of
PS coupling. However, for the PV coupling, only a modest increase 
in the effective mass is observed. These modified dispersion relations due 
to the presence of the external field can have substantial influence in 
the phenomenological aspect of the mesons both in the context of neutron stars as well as relativistic heavy ion collisions.
\end{abstract}
\maketitle

\section{Introduction}

The study of the properties of strongly interacting matter in magnetic field has become a research topic of contemporary interest \cite{Tuchin:2010gx,Tuchin:2013bda,Tuchin:2014iua,Xia:2014wla,Gorbar:2013uga,Shovkovy:2012zn, Ayala:2015lta, Ayala:2015qwa, Ayala:2014uua, Ayala:2014gwa, Ayala:2014mla, loewe2014}. The applicability of field theoretical calculations with the introduction of magnetic field lies in the study of the phenomenology of compact stars which are laboratories of high density matter and fields with strengths as high as $eB\sim 1 MeV^2$ observed in some magnetars \cite {duncan92}. In fact, the incorporation of the magnetic field effects to color superconducting phases in the core of such stars can provide new insight into the physics of neutron stars \cite{mark,gorbar,qcdmag1,qcdmag2,fh,jorge,chinese,qcdmag3}. On the other hand, the other domain, namely Relativistic Heavy Ion Collisions (RHIC) can hardly be overlooked. Recently, it has been proposed that for off-central heavy ion collisions, the 
intensity of the 
magnetic field due to presence of charged species can be as high as $eB\sim m_{\pi}^2\sim 0.02 GeV^2$ (at RHIC) and $eB\sim 15 m_{\pi}^2\sim 0.3 GeV^2$ (at LHC) \cite{kharzeev08, skokov09}.

It will not be out of context here to recall that in view of comparable mass scales of mesonic matter with field strengths, the study of pions $(\pi ^0 $and $\pi^{\pm})$ begs further attention. Many authors have studied pion properties either restricted to the symmetric nuclear matter or performed calculations in the non-relativistic framework \cite{Gale:1987ki, Oset:1981ih, Migdal:1990vm, Dmitriev:1984ud, Henning:1993tf, Korpa:1995mj}. In a recent work, the authors of \cite{Biswas:2006ma, Biswas:2007zs} have shown pionic mode splitting in asymmetric nuclear matter (ANM). Using the approximation of the hard nucleon loop and suitable density expansions, they have studied pion propagation in matter in the framework of Chiral effective Lagrangian model \cite{Matsui:1982qc}. In \cite{Biswas:2007zs} they presented the density and asymmetry dependent pion dispersion relations and effective masses for the various charged states of pion considering both pseudoscalar and pseudovector representation of pion-nucleon 
interactions. It 
was shown that the effective pion masses had large values in the pseudoscalar representation compared to pseudovector representation. But none of them considered magnetic field effect in calculation of the self energies. In the context of heavy ion collisions, pions in nuclear matter might carry a bulk amount of entropy which is explained by a modified pion spectrum \cite{mishustin80}. The pion-nucleon physics has been further explored in the works of Anderson \cite{andersen12} where chiral perturbation theory has been used for the systematic calculation of the leading loop corrections to the thermal mass and decay constant of pions at finite temperature and magnetic background. Condensation of pions have been studied extensively in the works of \cite{migdal78, Colucci:2013zoa}. The magnetic field effects have been studied thoroughly in the context of NJL model \cite{klev,shovkovy+,gorbie,klim,hiller,boomsma2,chat}, PNJL model \cite{pnjlgat,pnjlkas}, quark meson model \cite{fraga1,frasca,rabbi,rashid,
Andersen:2012bq}, PQM model \cite{fragapol,skokov} and linear sigma model \cite{duarte}. In a 
recent work, the authors of \cite{machado14} showed the modification of the 
charged $B$ meson mass in presence of the external field. They have concluded that there is a substantial decrease of the mass of $B$ mesons in the limits of strong and weak magnetic field. In fact, this approximation of weak field has been further explored in the works of Ayala \textit{et. al.} \cite{Ayala:2015qwa}. In this work, the authors have found that the field assists in formation of gluon condensate and acts against quark de-confinement.

In view of these recent theoretical advancements, we re-visit the problem of pionic dispersion relations starting from a phenomenological Lagrangian. In our work, we will restrict ourselves to the calculation of the pion effective mass in uniform Gauge field to one loop order in vacuum. For our calculations, we introduce the Feynman propagators for a spin $1/2$ fermion in an external constant Abelian field, best described by the Schwinger's proper time formalism \cite{schwinger51}. This will manifest a consistent framework for treating mesonic matter under the influence of the weak limit of magnetic fields $(eB << m_{\pi}^2)$ compatible with strengths observed in the interior of neutron stars. Due to divergences inherited into the theory of self-energies, we will remove them by regularization and subsequent renormalisation (by counterterms) of the modified self energy of the pions. The study of the pionic dispersion 
relations in presence of matter will be reported in a 
future work\cite{inprep}.

The paper is organized as follows. In Section II, we discuss the formalism required for the explicit calculation of the pion self energies in presence of weak magnetic field. We perform the calculations for pseudoscalar and pseudovector coupling in subsections (A) and (B) respectively. We will illustrate a consistent formalism for the fermionic propagators using Schwinger's proper time approximation followed by regularization and renormalisation of the vacuum fluctuations. The results will be presented in Section III.
 Finally, in section(IV), we will summarize and explain possible phenomenological implication of our results.

\section{The pion self-energy}
The effective pion propagator is given by resumming the 
pion self-energy using the Dyson-Schwinger equation,
\be
D(q) = D^0(q) + D^0(q)\Pi(q)D(q),
\ee
where $D^0(q) = (q^2-m^2_{\pi}+i\epsilon)^{-1}$ is the bare propagator and 
$\Pi(q)$ is the pion self-energy. The effective propagator can be written as,
\be 
D(q)=\frac{1}{[D^0(q)]^{-1}-\Pi(q)}\label{effpro}
\ee
The pole of the effective propagator determines the dispersion relation of the 
system with the modified mass $m_{\pi}^*=\sqrt{m^2_{\pi}+\Pi(m_{\pi}^*,{\bf q}=0)}$. 
To include the effect of external magnetic field, we use Schwinger's proper-time 
method~\cite{schwinger51}.
Let us consider the magnetic field along the $z$ direction with the choice of vector potential $\vec A=(-By/2, Bx/2, 0)$. In this choice, the momentum- space Schwinger propagator can be written as \cite{schwinger51},
\be
S(k) = \int_0^{\infty} \frac{ds}{\cos(eBs)}\exp\Bigg[is(k^2_{||}
-k^2_{\perp}\frac{\tan(eBs)}{eBs}-m^2+i\epsilon)\Bigg]
\big[(m+k\!\!\!/_{||})\exp(ieBs\sigma_3)-\frac{k\!\!\!/_{\perp}}{\cos(eBs)}\big].\label{Schwinger}
\ee
where $m$ is the mass of the fermion.
Note that the Schwinger's propagator for charged fermions usually
contains a phase factor. However, by suitable gauge transformation of the
vector potential, the phase factor can be removed and we can work with the momentum
representation of the Fermion propagators \cite{Ayala:2014uua,Ayala:2015qwa}.
We decompose the metric tensor into two parts $g^{\mu\nu} =g^{\mu\nu}_{||}-g^{\mu\nu}_{\perp}$, 
where $g^{\mu\nu}_{||} = {\rm diag}(1,0,0,-1)$ and $g^{\mu\nu}_{\perp} = {\rm diag}(0,1,1,0)$.
In this  notation we use $k^2_{||} = k_0^2-k^2_z$, $k^2_{\perp} = k^2_x+k^2_y$ and $\sigma_3 = i\gamma_1\gamma_2$. 
As we focus on the weak field approximation of the propagators $eB<< m_\pi^2$, the propagators can be recast up to order $(eB)^2$ as \cite{Chyi:1999fc, Navarro:2010eu},
\be
S(k) = S^{(0)}(k)+eB\, S^{(1)}(k) + (eB)^2\, S^{(2)}(k) + {\mathcal O}((eB)^3) 
\ee
with 
\be
S^{(0)}(k) = \frac{k\!\!\!/ + m}{k^2-m^2}
\ee
is the free fermionic propagator and 
\bea
S^{(1)}(k)  &=& \frac{i\gamma_1\gamma_2(\gamma.k_{||}+m)}{(k^2-m^2)^2}~~~~\\
S^{(2)}(k) &=& \frac{-2k^2_{\bot}}{(k^2-m^2)^4}[k\!\!\!/ + 
m-\frac{\gamma.k_{\bot}}{k_{\bot}^2}(k^2-m^2)]
\eea
are the weak field corrections to the propagator. 
\begin{figure}[htb]
\begin{center}
\includegraphics[scale=0.55,angle=0]{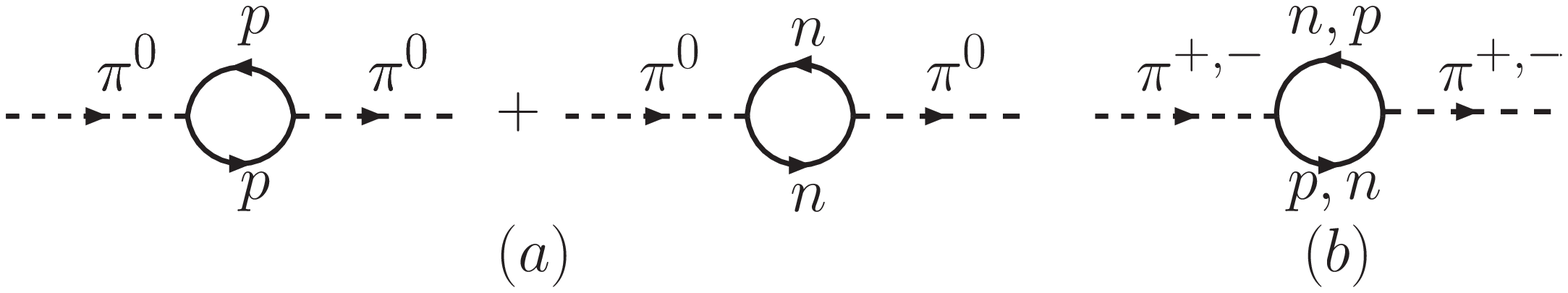}
\caption{(a) represents the one-loop self-energy diagram for $\pi^0$
and (b) represents the same for $\pi^{\pm}$.}
\label{fig0}
\end{center}
\end{figure}
\subsection{Dispersion relation with pseudoscalar $\pi$N coupling}
We use a phenomenological $PS$ pion-nucleon interaction to write the Lagrangian density as, 
\be
{\mathcal L}_{\rm int}^{\rm PS} =  -i{\rm g}_{\pi}{\bar \Psi}
\gamma_5({\vec \tau}\cdot {\vec \Phi_{\pi}})\Psi\label{lag1}
\ee
where ${\rm g}_{\pi}$ is the pion-nucleon coupling constant. In Eq.(~\ref{lag1}) 
$\Psi$ and $\Phi_{\pi}$ are the nucleon and pion fields respectively 
and $\vec \tau$ is the isospin operator. The one loop contribution
to the pion self-energy is given by,
\be
\Pi_{\pi}(q) = -i\int \frac{d^4k}{(2\pi)^4}{\rm Tr}[\{i\Gamma(q)\}iS_a(k)\{i\Gamma(-q)\}iS_b(k+q)]
\ee
where the subscripts $a$ and $b$ denote either $p$ (proton) or $n$ (neutron).
$\Gamma(q)$ is the corresponding vertex factor for 
pseudoscalar coupling of pions. Fig.(\ref{fig0}) will involve various
combinations of $n$ and $p$ depending 
upon the various charged states of pions.
Let us explicitly calculate the one loop self-energy for
$\pi^0$ shown in Fig.~\ref{fig0}(a). In the weak field limit, self-energy
for $\pi^0$ is expressed as:
\be
\Pi_{\pi^0}(q) = -i{\rm g}^2_{\pi}\int \frac{d^4k}{(2\pi)^4} 
{\rm Tr}[\gamma_5\,iS_p(k)\,\gamma_5\,iS_{\bar{p}}(k+q)] + [p\rightarrow n]
\label{pi0}
\ee
For neutral pion,  $\Gamma(q)= -i\gamma_5{\rm g}_{\pi}$ is the vertex 
factor.  
The non-vanishing contribution for $\Pi^0(q)$ can be written as 
\be
\Pi_{\pi^0}(q) = \Pi^{(0,0)}_{\pi^0}(q) + (eB)^2\,\Pi^{(1,1)}_{\pi^0}(q) 
+ (eB)^2\,\Pi^{(2,0)}_{\pi^0}(q) + (eB)^2\,\Pi^{(0,2)}_{\pi^0}(q)
\ee
Upon evaluation of Dirac traces, it is seen that all terms proportional to $(eB)$ 
have vanishing traces. The reason for the cancellation is either due to odd number of $\gamma$ matrices or off diagonal elements of the metric tensor. 
The $B-$ independent vacuum contribution of the self energy is expressed as,
\bea
\Pi^{(0,0)}_{\pi^0}(q) &=& -i{\rm g}^2_{\pi}\int \frac{d^4k}{(2\pi)^4} 
{\rm Tr}[\gamma_5\,iS^{(0)}_p(k)\,\gamma_5\,iS^{(0)}_p(k+q)] + [p\rightarrow n]\nn\\
&=& i{\rm g}^2_{\pi}\int \frac{d^4k}{(2\pi)^4} 
{\rm Tr}[\gamma_5(k\!\!\!/+m_p)\gamma_5(k\!\!\!/+q\!\!\!/+m_p)]
\frac{1}{[k^2-m_p^2][(k+q)^2-m_p^2]} + [p\rightarrow n],\nn\\
&=&i{\rm g}^2_{\pi}\int \frac{d^4k}{(2\pi)^4}\,4\big[m^2_p-k.(k+q)\big]
\frac{1}{(k^2-m_{p}^2)((k+q)^2-m_{p}^2)} + [p\rightarrow n]\nn\\
&=&\frac{{\rm g}^2_{\pi}}{4\pi^2}\Big[\frac{q^2}{3}+
\big[1+\frac{1}{\varepsilon}-\gamma_E+\log(4\pi\mu^2)\big](m_p^2-\frac{q^2}{2})
-\int^1_0 dx\big(m_p^2-3x(1-x)q^2\big)\log[m_p^2-x(1-x)q^2]\Big]+ [p\rightarrow n],\nn\\
\label{pi00}
\eea
where $m_p$  $(m_n)$ is the mass of proton (neutron).
Here, $\varepsilon=2-\frac{N}{2}$ and $\mu$ is an arbitrary scale parameter.
$\gamma_E$ is the Euler-Mascheroni constant. It is clearly seen that $\varepsilon$
in Eq.(\ref{pi00}) contains the singularity and it diverges as $N\rightarrow4$.
To remove the divergences, we need to add the counterterms in 
the Lagrangian~\cite{Matsui:1982qc}. 
Hence, after performing the renormalization one can obtain the modified 
pion self energy as (explicitly shown in the Appendix A), 
\bea
\Pi^{(0,0)}_{\pi^0 \bf{R}}&=& - \frac{g^2_{pp\pi^0}}{4\pi^2}\int^1_0dx \Bigg[
\frac{(q^2-m_{\pi^0}^2)\,x(1-x)\big[m_p^2-3m_{\pi}^2x(1-x)\big]}{m_p^2-m_{\pi^0}^2x(1-x)}\nn\\
&+&\big[m_p^2-3q^2\,x(1-x)\big]
\log\frac{\Delta_R}{m_p^2-m_{\pi^0}^2\,x(1-x)}
\Bigg]+ [p\rightarrow n],
\label{renormvacPS}
\eea
where $\Delta_R =m_p^2-q^2x(1-x)$ and $m_{\pi^0}$ is the mass of the neutral pion.
 The first term of the external magnetic field dependent contribution to the self energy can be obtained as (see Appendix B),
\bea
\Pi^{(1,1)}_{\pi^0}(q) &=& -i{\rm g}^2_{\pi}\int \frac{d^4k}{(2\pi)^4} 
{\rm Tr}[\gamma_5\,iS^{(1)}_p(k)\,\gamma_5\,iS^{(1)}_p(k+q)]\nn\\
&=& -\frac{{\rm g}^2_{\pi}}{4\pi^2}\int_0^1 dx\,x(1-x)
\Big[\frac{1}{\Delta_R}+\frac{m_p^2+x(1-x)q^2_{||}}{\Delta_R^2}\Big]\label{pi11}
\eea
The term proportional to $(eB)^2$ \rm {i.e.} $\Pi^{(2,0)}_{\pi^0}(q)$ is given by (see Appendix B for details),
\bea
\Pi^{(2,0)}_{\pi^0}(q) &=& -i{\rm g}^2_{\pi}\int \frac{d^4k}{(2\pi)^4} 
{\rm Tr}[\gamma_5\,iS^{(2)}_p(k)\,\gamma_5\,iS^{(0)}_p(k+q)]\nn\\
&=& -\frac{{\rm g}^2_{\pi}}{4\pi^2}
\Bigg[\int_0^1 dx(1-x)^3\big[\frac{1}{\Delta_R}+
\frac{q^2 x(1-x)+q^2_{\bot}x(4x-1)+m_p^2}{3\Delta_R^2}
+\frac{2x^2q_{\bot}^2[q^2x(1-x)+m_p^2]}{3\Delta_R^3}\big]\nn\\ 
&&~~~~~~+\int_0^1 dx(1-x)^2\big[\frac{1}{\Delta_R}
-\frac{q^2_{\bot}x(1-x)}{\Delta_R^2}\big]\Bigg]\label{pi200}
\label{pi20}
\eea
In order to obtain the contribution of $\Pi^{(0,2)}_{\pi^0}(q)$, we
just replace $k\leftrightarrow (k+q)$ in Eq.(\ref{pi20}) and we see that 
$\Pi^{(0,2)}_{\pi^0}(q)$ and $\Pi^{(2,0)}_{\pi^0}$ are identical. It is 
seen that the contribution of the magnetic field dependent 
self-energy is finite, \textit{i.e.} no divergences appear. To simplify our calculation,
we consider the proton and neutron have the same mass denoted by $m$. The complete expression of the self-energy for 
$\Pi_{\pi^0}(q)$ for the pseudoscalar coupling can be written as, 
\bea
\Pi_{\pi^0}(q) &=& -\frac{g^2_{\pi}}{2\pi^2}\Bigg[\int^1_0dx \Big[
\frac{(q^2-m_{\pi^0}^2)\,x(1-x)(m^2-3q^2x(1-x))}{\Delta_R}+(m^2-3m_{\pi^0}^2\,x(1-x))
\log\frac{\Delta_R}{m^2-m_{\pi^0}^2\,x(1-x)}\Big]\nn\\
&+&\frac{(eB)^2}{2}\int^1_0dx\,x(1-x)\Big(\frac{1}{\Delta_R}+
\frac{m^2+x(1-x)q^2_{||}}{\Delta_R^2}\Big)\nn\\
&+& (eB)^2\Bigg\{\int_0^1 dx\,(1-x)^3\big[\frac{1}{\Delta_R}+
\frac{q^2 x(1-x)+q^2_{\bot}x(4x-1)+m^2}{3\Delta_R^2}
+\frac{2x^2q_{\bot}^2[q^2x(1-x)+m^2]}{3\Delta_R^3}\big]\nn\\ 
&& ~~~~~~~~+\int_0^1 dx(1-x)^2\big[\frac{1}{\Delta_R}
-\frac{q^2_{\bot}x(1-x)}{\Delta_R^2}\big]\Bigg\}\Bigg].\label{pipi0}
\eea
 If we do not distinguish between neutron and proton mass, the expression for 
the self-energy of $\Pi_{\pi^+}(q)$ and $\Pi_{\pi^-}(q)$ are identical. For 
$\pi^{\pm}$ the coupling constant ${\rm g}_{\pi}$ gets replaced by 
$\sqrt{2}{\rm g}_{\pi}$. Using the previous procedure we can easily calculate 
the self-energy for $\Pi_{\pi^{\pm}}(q)$. The contribution from the diagram 
in Fig.~\ref{fig0}(b) for $\Pi_{\pi^{\pm}}(q)$ is obtained as follows,
\bea
\Pi_{\pi^\pm}(q) &=& -\frac{g^2_{\pi}}{2\pi^2}\Bigg[
\int^1_0dx \Big[
\frac{(q^2-m_{\pi^{\pm}}^2)\,x(1-x)(m^2-3q^2x(1-x))}{\Delta_R}+(m^2-3m_{\pi^{\pm}}^2\,x(1-x))
\log\frac{\Delta_R}{m^2-m_{\pi^{\pm}}^2\,x(1-x)}\Big]\nn\\
&+& (eB)^2\Bigg\{ \int_0^1 dx(1-x)^3\big[\frac{1}{\Delta_R}+\frac{q^2 x(1-x)+q^2_{\bot}x(4x-1)+m^2}{3\Delta_R^2}
+\frac{2x^2q_{\bot}^2[q^2x(1-x)+m^2]}{3\Delta_R^3}\big]\nn\\ 
&&~~~~~~~+ \int_0^1 dx(1-x)^2\big[\frac{1}{\Delta_R}
-\frac{q^2_{\bot}x(1-x)}{\Delta_R^2}\big]\Bigg\}\Bigg].\label{pipipm}
\eea

 Using Eq.(\ref{effpro}) and Eqs.(\ref{pipi0}),(\ref{pipipm}) we can 
 calculate the dispersion relations for $\pi^0$ and $\pi^{\pm}$ for pseudoscalar 
 coupling. Considering a relativistic particle with mass $m_\pi$ moving in a homogeneous external magnetic field $\vec{B}$, directed along the $z$ axis, the modified pion mass ($m^{*\,2}_{\pi}$) is given by the following expression,
 \be
 m^{*\,2}_{\pi} = m^2_{\pi} +{\rm Re}\,\Pi(m^{*\,2}_{\pi},{\bf q}=0,B).\label{effmass}
 \ee
The above expression is obtained by defining the effective pion masses by the positions of the pole of the propagator.

\subsection{Dispersion relation with pseudovector $\pi$N coupling}
To obtain the self-energy in pseudovector interaction, we start from the interaction  
Lagrangian,
\be
\mathcal{L}_{\rm int}^{\rm PV} = -\frac{f_{\pi}}{m_{\pi}}{\bar\Psi}^{\prime}\gamma_5\gamma^{\mu}
\partial_{\mu}({\bf{\tau}} .\Phi_{\pi}^{\prime})\Psi^{\prime}
\ee
where $f_{\pi}$ is the pseudovector coupling constant. The vertex factor for 
pseudovector coupling is $(-i)\frac{f_{\pi}}{m_{\pi}}\gamma_5q\!\!\!/$.
First we discuss about the neutral pion self-energy. 
The contribution for the field free part of $\pi^0$ is
\bea
\Pi^{(0,0)}_{\pi^0}(q) &=& -i\,(\frac{f_{\pi}}{m_{\pi}})^2
{\rm Tr}\Bigg[\gamma_5q\!\!\!/S^{(0)}_p(k)\gamma_5q\!\!\!/S_p^{(0)}(k+q)\Bigg]+ [p\rightarrow n]\nn\\
&=& (\frac{f_m}{m_{\pi}})^2\frac{q^2}{4\pi^2}
\int_0^1dx\,2m^2_p\Big[-\frac{1}{\varepsilon}+\gamma_E+\ln(\frac{\Delta_R}{4\pi\mu^2})\Big] + [p\rightarrow n]
\label{PV00}
\eea
It is seen that Eq.(\ref{PV00}) diverges as $N\rightarrow4$.
To remove the divergences, we use simple subtraction obtaining,
\bea
\Pi^{(0,0)}_{{\pi^0}_{{\bf R}}}(q) &=&\Pi^{(0,0)}_{\pi^0}(q)-\Pi^{(0,0)}_{\pi^0}(m_{\pi})\nn\\
&=&(\frac{f_m}{m_{\pi}})^2\frac{q^2}{4\pi^2}
\int_0^1dx\,2m^2\Big[\ln(\frac{m^2-q^2x(1-x)}{m^2-m_{\pi^0}^2x(1-x)})\Big] + [p\rightarrow n]\\
&=&(\frac{f_m}{m_{\pi}})^2\frac{q^2}{4\pi^2} 4m^2
\Bigg[\frac{\sqrt{4m^2-q^2}}{q}\tan^{-1}\big(\frac{q}{\sqrt{4m^2-q^2}}\big)
-\frac{\sqrt{4m^2-m^2_{\pi^0}}}{m_{\pi^0}}\tan^{-1}\frac{m_{\pi^0}}{\sqrt{4m^2-m^2_{\pi^0}}}
\Bigg]
\label{PV00R}
\eea
In case of pseudovector coupling we do not find any linear order contribution of $(eB)$
similar to the case of pseudoscalar coupling for the same reason as mentioned before. Moreover, the contribution 
of magnetic field comes from ${\mathcal O}((eB)^2)$ terms. Hence, the value of 
$\Pi^{(1,1)}_{\pi^0}(q)$ is given by (See Appendix C), 
\bea
 \Pi^{(1,1)}_{\pi^0}(q) &=&  -i\,(\frac{f_{\pi}}{m_{\pi}})^2\int \frac{d^4k}{(2\pi)^4}
{\rm Tr}\Bigg[\gamma_5q\!\!\!/S_p^{(1)}(k)\gamma_5q\!\!\!/S_p^{(1)}(k+q)\Bigg] \nn\\
&=& (\frac{f_{\pi}}{m_{\pi}})^2\frac{1}{4\pi^2}\int_0^1x\,(1-x)\,dx\,
\Big[\frac{x(x-1)q^2(2q_{||}^2-q^2)-m^2q^2_{||}}{\Delta_R^2}\Big].\label{PV11}
\eea  
The expression for $\Pi^{(2,0)}_{\pi^0}(q)$ is as follows (See Appendix C),
\bea
\Pi^{(2,0)}_{\pi^0}(q) = &=& -i\,(\frac{f_{\pi}}{m_{\pi}})^2 \frac{d^4k}{(2\pi)^4}
{\rm Tr}\Big[\gamma_5q\!\!\!/S_p^{(2)}(k)\gamma_5q\!\!\!/S_p^{(0)}(k+q)\Big]\nn\\
&=& (\frac{f_{\pi}}{m_{\pi}})^2\frac{1}{4\pi^2}
\Bigg[\int_0^1dx\frac{(1-x)^3}{3}\,q^2\,\Big[\frac{3}{2\Delta_R}+
x(1-x)\frac{q^2_{\perp}}{\Delta_R^2}-[x(1-x)q^2+m^2]
(\frac{1}{\Delta_R^2}+2x^2\frac{q^2_{\perp}}{\Delta_R^3})\Big]\nn\\
&&~~~~~~~~~~~~~+\int_0^1dx(1-x)^2\Big[\frac{q^2+q^2_{\perp}}{\Delta_R}+
x(1-x)\frac{q^2q^2_{\perp}}{\Delta_R^2}\Big]\Bigg]\label{PV20}. 
\eea
The value of $\Pi^{(0,2)}_{\pi^0}(q)$ is identical with $\Pi^{(2,0)}_{\pi^0}(q)$ 
because we consider $m_p=m_n=m$. Now we can easily calculate the self-energy 
for neutral and charged pions and the expressions are as follows
\bea
\Pi_{\pi^0}(q) &=& \Pi^{(0,0)}_{{\pi^0}_{{\bf R}}}(q)
+ (eB)^2\Big[\Pi^{(1,1)}_{\pi^0}(q)+2\,\Pi^{(2,0)}_{\pi^0}(q)\Big]\nn\\
\eea

Similarly, following the same procedures, we can obtain the charged pion self energies for the pseudovector coupling:
\bea
\Pi_{\pi^{\pm}}(q) &=& (\frac{f_m}{m_{\pi}})^2\frac{1}{2\pi^2}\Bigg[
2q^2m^2\Bigg\{\frac{\sqrt{4m^2-q^2}}{q}\tan^{-1}\big(\frac{q}{\sqrt{4m^2-q^2}}\big)
-\frac{\sqrt{4m^2-m^2_{\pi^{\pm}}}}{m_{\pi^{\pm}}}\tan^{-1}
\frac{m_{\pi^{\pm}}}{\sqrt{4m^2-m^2_{\pi^{\pm}}}}\Bigg\}\nn\\
&&~~~~~~+(eB)^2\Bigg\{\int_0^1dx\frac{(1-x)^3}{3}\,q^2\,\Big[\frac{3}{2\Delta_R}+
x(1-x)\frac{q^2_{\perp}}{\Delta_R^2}-[x(1-x)q^2+m^2]
(\frac{1}{\Delta_R^2}+2x^2\frac{q^2_{\perp}}{\Delta_R^3})\Big]\nn\\
&&~~~~~~~~~~~~~+\int_0^1dx(1-x)^2\Big[\frac{q^2+q^2_{\perp}}{\Delta_R}+
x(1-x)\frac{q^2q^2_{\perp}}{\Delta_R^2}\Big]\Bigg\}\Bigg]\label{PVpipm}.
\eea




%

\begin{figure}[h]
 \includegraphics[height=6cm, angle=0]{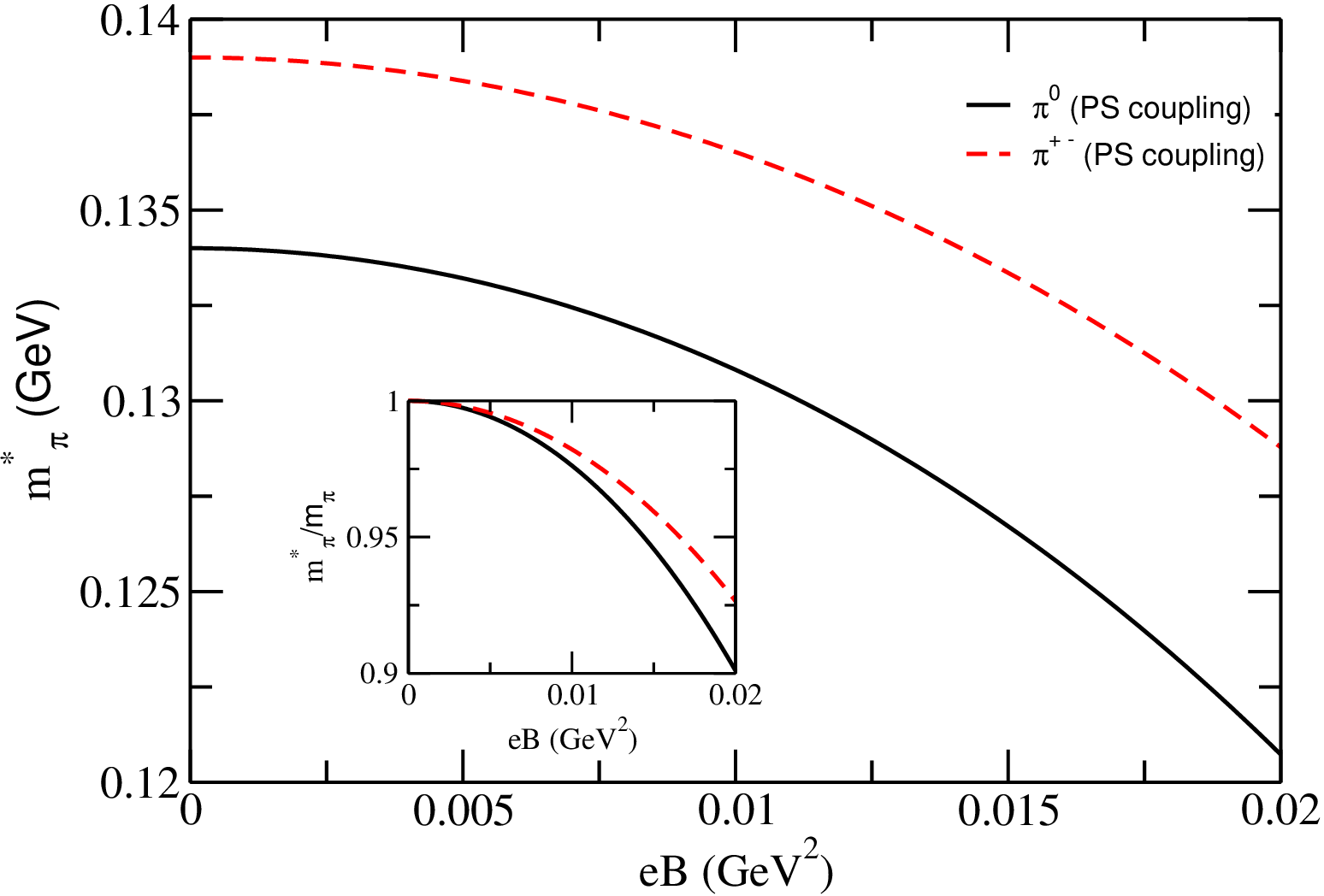}
\caption{ (Color online) Effective pion mass as a function of the magnetic field for PS coupling.}\label{fig1}
\end{figure}

\begin{figure}
 \includegraphics[height=5cm, angle=0]{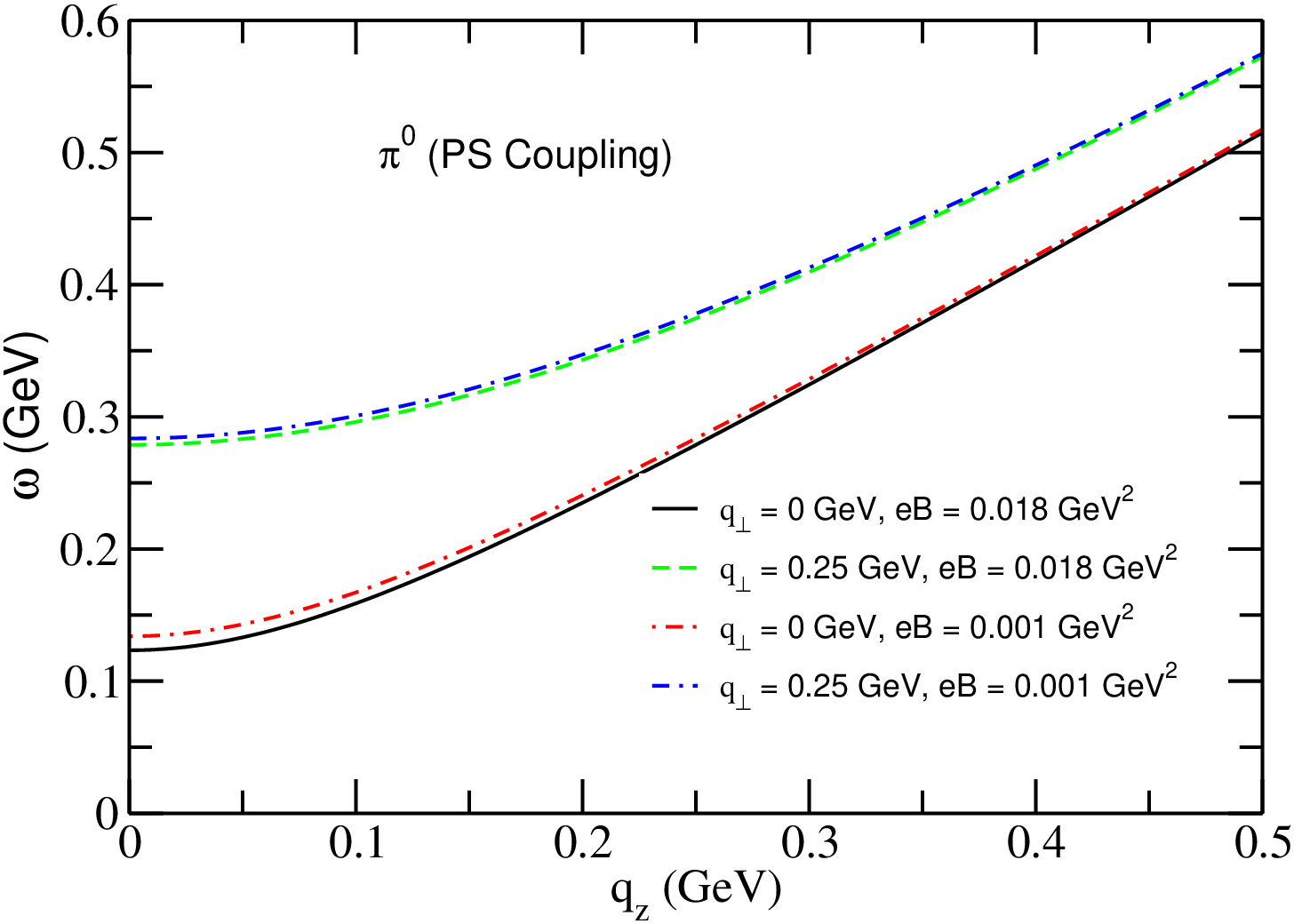}~~~~
 \includegraphics[height=5cm, angle=0]{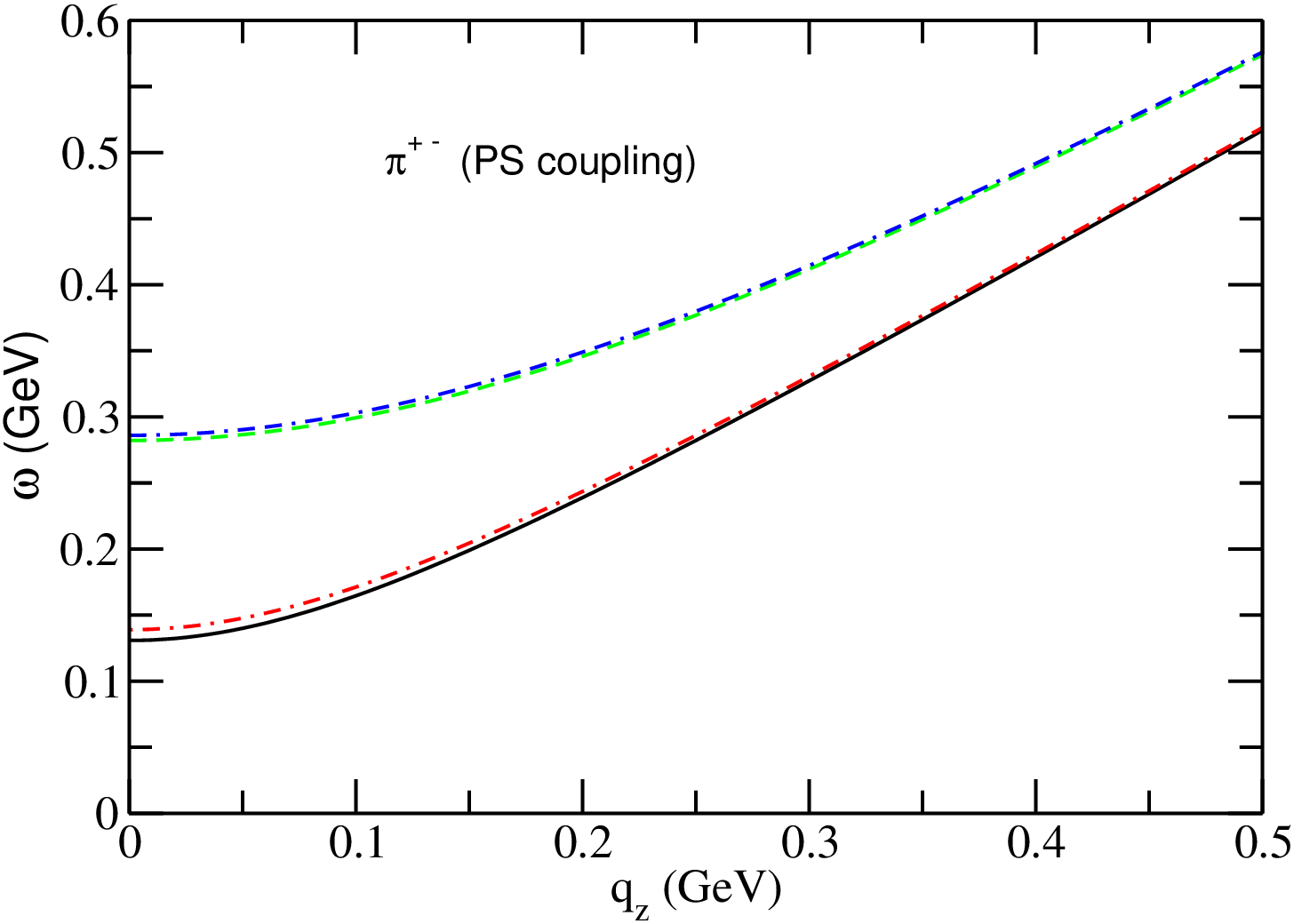}
\caption{(Color online) The left panel shows the dispersion relation for neutral pion with the z- component of momentum for pseudoscalar coupling. The right panel shows a similar plot for the case of charged pions}\label{fig2}
\end{figure}

\begin{figure}[h]
 \includegraphics[height=6cm, angle=0]{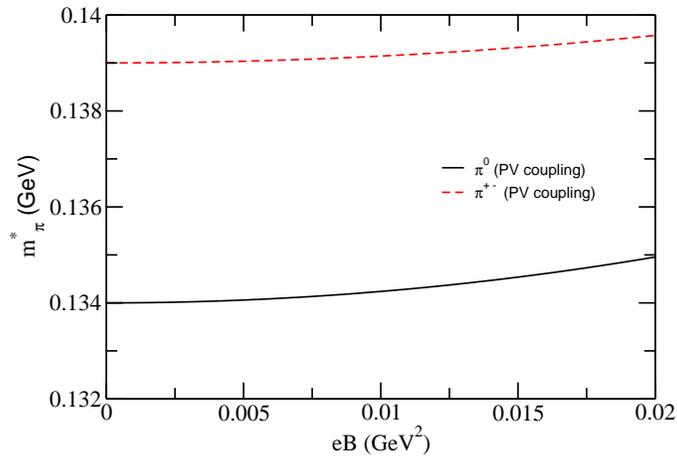}
\caption{ (Color online) Effective pion mass as a function of the magnetic field for PV coupling.}\label{fig3}
\end{figure}

\begin{figure}
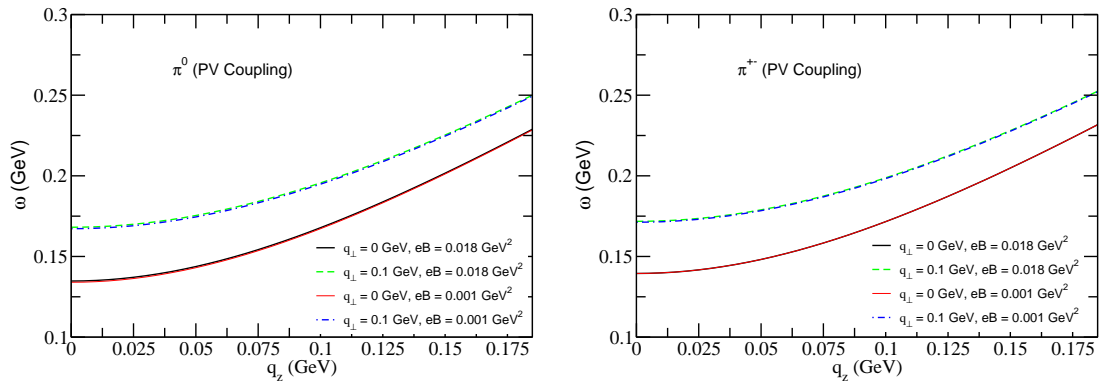

 \includegraphics[height=5cm, angle=0]{PV_pip0_disp.eps}~~~~
 \includegraphics[height=5cm, angle=0]{PV_pipm_disp.eps}
\caption{(Color online) The left panel shows the dispersion relation for neutral pion with the z- component of momentum for pseudovector coupling. The right panel shows a similar plot for the case of charged pions.}\label{fig4}
\end{figure}

\section{Results and discussions}
We have considered the case of weak magnetic field for the numerical evaluation of our results. We have used the condition that the field strength is much lower than the squared pion mass $(eB)\ll m_{\pi}^2$ which is our consideration for the weak field limit of the field. The pseudoscalar coupling $g_\pi^2/(4\pi)$ is $14.1$ \cite{Piekarewicz:1993ad} and for the pseudovector case, we have taken the coupling $f_\pi^2/(4\pi)$ as $0.08$ \cite{ericssonbook}. Here, we have used the relation $g^2/(2 m_p)^2= (f/m_\pi)^2$. Quantum fluctuations will invoke the possibility of the usage of the Schwinger's proper time propagators in presence of external magnetic field. Thus, we obtain the vacuum contribution which needs to be suitably renormalized to eliminate the divergences in the theory. We have adopted the method of addition of mass counterterms for the PS coupling and method of subtraction of terms for the PV coupling respectively. The details about the renormalization procedure for PS coupling is highlighted in the Appendix (A). 
The correction terms which are quadratic in field parameter $(eB)$ contributes to the self energy of pions at one loop order and are devoid of any divergences. The modified pion masses $m_{\pi}^{*0}$ and $m_{\pi}^{*\pm}$ are defined by the position of the pole of the effective propagator and setting the three momentum label equal to zero. Further, the dispersion relation of the pions are evaluated numerically by setting the $z-$ component of momentum up to a maximum value of $0.175$ GeV. The bare pion masses have been fixed as $m_{\pi}^{0}=0.134$ GeV and $m_{\pi}^{\pm}=0.139$ GeV. 

At first, we will consider the case of PS coupling of pions.
Fig.(\ref{fig1}) displays the effective mass of pion $(m_\pi^*)$ as a function of the background magnetic field and shows a significant decrease in the effective mass for both the neutral as well as charged pions in PS coupling. 
In addition, it is seen in the inset of Fig.(\ref{fig1}) that the rate of decrease of the pion mass with the magnetic field is greater for the neutral pion than the charged pion.

In Fig.(\ref{fig2}), we present the graph of the dispersion relation of neutral pion (left panel) and charged pion (right panel) for PS coupling scheme. In both the graphs, the trend is similar in nature. In the limit of low $z- component$ of momentum, there is a moderate increase of the energy variable for different perpendicular components of the momentum $(0$ $GeV$ and $0.25$ $GeV)$ coupled with field strengths $(0.018$ $GeV^2$ and $0.001 GeV^2)$.

In contrast to PS coupling, it is observed that for PV coupling the effective pion masses increases slightly with the increase of external magnetic field strength as shown in Fig.(\ref{fig3}). The dispersions of neutral pion (left panel) and charged pions (right panel) are displayed in Fig.(\ref{fig4}) at different magnetic field strengths and perpendicular component of momentum ($q_\perp$). We observe the increasing nature of pion energy similar to that in case of PS coupling. It is noticed that for a fixed $q_{\perp}$ the dispersion relations are insensitive to the magnetic field both for PS and PV couplings. 

\section{Summary and Conclusion}
In this work, we have re-visited the modification of the pion dispersion relations by the introduction of the external magnetic field on the charged and neutral pions. For our purpose, we have used Schwinger's proper time method of fermion propagator in presence of background magnetic field to effectively describe proton propagators in the one loop self energy corrections. The effect of the external magnetic field appears as corrections of order $(eB)^2$ over the vacuum contribution to the pion self energy which are relevant for the study of neutron stars and relativistic heavy ion collisions.
The phenomenology of pions in nuclear matter is generally described by a
chiral invariant pion-nucleon interaction which leads to the additional 
Lagrangian term $\mathcal{L}_{\pi NN}=-(g_{\pi NN}/2g_A m_{N})^2
\bar\Psi_N\gamma^\mu\tau\Psi_N(\vec\Pi\times\del_{\mu}\vec\Pi)$ known
 as the Weinberg-Tomozawa term in the literature. We have calculated the contribution 
from this term to the pion self energy. We have found that the contribution
for the corresponding diagram for the $\pi \pi NN$ interaction vanishes at the
$(eB)$ and $(eB)^2$ order of the external magnetic field.
From the numerical estimates, we conclude decreasing nature of effective pion masses in case of $PS$ coupling while an increasing nature is noticed for $PV$ coupling.
We re-confirm the result of the vacuum fluctuation free field contribution to the pion self energy with an earlier work \cite{Piekarewicz:1993ad, Piekarewicz:1993ff}. The 
values of the weak magnetic field is considered up to $0.018 m_\pi^2$ as relevant in the phenomenological scenario. However, we do not include the medium modifications in our calculation which will be reported soon in a future work. The results obtained here serve as a theoretical framework for study at finite density and/ or temperature in presence of arbitrary magnetic field \cite{inprep}.
Finally, it should be noted that we have not incorporated the nucleon's magnetic moment in the present work. Inclusion of this will
contribute in $(eB)$ order. However, we intend to investigate this in future.
\section*{Appendix A: Renormalization of vacuum fluctuation of pion self energy}
As the free space contribution of the pion self energy is plagued with infrared divergences, we need to add counterterms with the original interaction Lagrangian to make it finite. Therefore, the counterterm Lagrangian is written as \cite{Matsui:1982qc},
\bea
{\mathcal L}_{CT}=-\f{1}{2!}\beta_1\Phi_{\pi}.(\del^2+m_\pi^2).\Phi_\pi+\f{1}{2!}\beta_2\Phi_{\pi}^2
\label{countL}
\eea
The values of the counterterms $\beta_1$ and $\beta_2$ are determined from the renormalization conditions as,
\bea
\beta_1&=&\Big(\f{\del\Pi(q)}{\del q^2}\Big)_{q^2=m_\pi^2}\nn\\
\beta_2&=&(\Pi)_{q^2=m_\pi^2}
\eea
We have obtained the following counterterms for renormalisation of the vaccuum fluctuation part,
 \bea
\beta_1 &=& -\f{g^2_\p}{4\p^2} \Bigg[ \f{1}{3}  -\f{1}{2}
\big(1+\f{1}{\ep}-\g+\ln(4\p\m^2) \big) \nn \\
&+&   \int^1_0 dx~3x(1-x)\ln( m^2-m^2_\p x(1-x))\nn \\
&+&    \int^1_0 dx  \f{m^2x(1-x)-3m^2_\p
x^2(1-x)^2}{m^2-m^2_\p x(1-x)}  \Bigg]+ [p\rightarrow n]   
\eea
\bea
\beta_2 &=& -\f{g^2_\p}{4\p^2}\Bigg[ \f{m^2_\p}{3}+\lt(m^2-\f{m^2_\p}{2}\rt)
\big(1+\f{1}{\ep}-\g +\ln(4\p\m^2)\big)  \nn \\
&-&   \int^1_0 dx~\lt(m^2-3m^2_\p x(1-x) \rt) \ln(m^2-m^2_\p x(1-x))\Bigg]+ [p\rightarrow n]
\eea
These two counterterms ensure that the pion propagators $D_\pi=[q^2-m_\pi^2-\Pi_{R}(q)]^{-1}$ reproduces the physical mass of pions in free space. Thus the renormalized pion self- energy for PS coupling is obtained as,
\bea
\Pi_{\bf R}(q, m_\pi)=\Pi(q)-\beta_1(q^2-m_\pi^2)-\beta_2
\eea
which we have used in the calculation of Eq.(\ref{renormvacPS}).
\section*{Appendix B: Details of calculation of self energy for pseudo- scalar coupling}
In this Appendix, we highlight some of the important steps required for calculation of the pion self- energy using the Lagrangian for PS coupling. We start by calculation of the traces of the terms in the self- energy as,
\bea
&&{\rm Tr}\Big[\gamma_5S_a(k)\gamma_5S_b(k+q)\Big]\nn\\
&=&{\rm Tr}\Bigg[\gamma_5S^{(0)}_a(k)\gamma_5qS_b^{(0)}(k+q)
+(e B)\Big\{\gamma_5S^{(0)}_a(k)\gamma_5S_b^{(1)}(k+q)
+\gamma_5S^{(1)}_a(k)\gamma_5S_b^{(0)}(k+q)\Big\}\nn\\
&+&(e B)^2\Big\{\gamma_5S^{(1)}_a(k)\gamma_5S_b^{(1)}(k+q)+
\gamma_5S^{(0)}_a(k)\gamma_5S_b^{(2)}(k+q)+
\gamma_5S^{(2)}_a(k)\gamma_5S_b^{(0)}(k+q)\Big\}+\mathcal{O}((e B)^3)\Bigg]\nn\\
&\simeq& T_1 + (e B)[T_2 + T_3] + (e B)^2[T_4 + T_5 + T_6]
\eea  
where,
\bea
T_1 &=& {\rm Tr}[\gamma_5S^{(0)}_a(k)\gamma_5S_b^{(0)}(k+q)]\nn\\
T_2 &=& {\rm Tr}[\gamma_5S^{(0)}_a(k)\gamma_5S_b^{(1)}(k+q)]\nn\\
T_3 &=& {\rm Tr}[\gamma_5S^{(1)}_a(k)\gamma_5S_b^{(0)}(k+q)]\nn\\
T_4 &=& {\rm Tr}[\gamma_5S^{(1)}_a(k)\gamma_5S_b^{(1)}(k+q)]\nn\\
T_5 &=& {\rm Tr}[\gamma_5S^{(0)}_a(k)\gamma_5S_b^{(2)}(k+q)]\nn\\
T_6 &=& {\rm Tr}[\gamma_5S^{(2)}_a(k)\gamma_5S_b^{(0)}(k+q)]
\eea
$T_1$ gives the field free case for the pion self energy. As mentioned previously in the text, $T_2$ and $T_3$ are individually zero as the traces vanish. The terms $T_4, T_5$ and $T_6$ contribute to the first order in field corrections which are quadratic in the magnetic field variable. Using the standard procedure of Feynman parametrization (FP) \cite{Peskin:1995ev}, the non- renormalized vacuum contribution to the self energy is obtained as Eq.(\ref{pi00}). Further, renormalization of Eq.(\ref{pi00}) gives Eq.(\ref{renormvacPS}) devoid of any divergences.
$T_4$ is evaluated as,
\bea
T_4 = {\rm Tr}[\gamma_5S^{(1)}_a(k)\gamma_5S_b^{(1)}(k+q)]
\eea
As we are considering the proton loop in Fig.(\ref{fig0}(a)), both the masses are to be considered as proton mass. Therefore, we re-write the above equation as,
\bea
 T_4 &=& {\rm Tr}[\gamma_5\gamma_1\gamma_2(\gamma.k_{||}+m_p)\gamma_5\gamma_1\gamma_2
(\gamma.(k+q)_{||}+m_p)\frac{1}{[k^2-m_p^2]^2[(k+q)^2-m_p^2]^2}\nn\\
&=& 4\big[m^2_p-k_{||}.(k_{||}+q_{||})\big]
\frac{1}{[k^2-m_p^2]^2[(k+q)^2-m_p^2]^2}\nn\\
\eea
where we have used,
\bea
\gamma_1\gamma_2(\gamma.k_{||})&=&(\gamma.(k+q)_{||})\gamma_1\gamma_2\nn\\
\gamma_1\gamma_2q\!\!\!/\gamma_1\gamma_2&=&2(\gamma.q)_{\perp}-q\!\!\!/
\eea

Using FP (and substituting $k\rightarrow (k-xq)$), we obtain,
\bea
T_4=\f{4}{[k^2_{||}-\Delta_{||}]^4}\Bigg[m_p^2+x(1-x)q_{||}^2-k_{||}^2+(2x-1)(k_{||}.q_{||})\Bigg]
\eea
where, $\Delta_{||} = k^2_{\perp}+\Delta_R$.
Solution of the momentum integral is as follows,
\bea
&&\int \frac{d^4k}{(2\pi)^4} T_4 \nn\\
&=&4\int_0^1x(1-x)\,dx\Gamma[4]\int \frac{d^4k}{(2\pi)^4}\f{m_p^2+x(1-x)q_{||}^2-k_{||}^2+(2x-1)(k_{||}.q_{||})}{[k^2_{||}-\Delta_{||}]^4}
\eea
In order to evaluate the above integration, we will use 4-momentum integrals involving the parallel and perpendicular components. Here, we list the following identities that will be required for the evaluation of the pion self energy.

Identity 1:
\bea
&&\int \frac{d^4k}{(2\pi)^4}\frac{1}{[k^2_{||}-\Delta_{||}]^4} = 
i\,\int \frac{d^2k_{\perp}}{(2\pi)^2}\int \frac{d^2k_{E_{||}}}{(2\pi)^2}
\frac{1}{[k^2_{E_{||}}+\Delta_{||}]^4}\nn\\
&=& i\,\int \frac{d^2k_{\perp}}{(2\pi)^2}\frac{1}{(4\pi)}
\frac{\Gamma[4-1]}{\Gamma[4]}(\frac{1}{\Delta_{||}})^{4-1} = i\, \frac{1}{(4\pi)^2}\frac{\Gamma[3]}{\Gamma[4]}
\frac{\Gamma[3-1]}{\Gamma[3]}(\frac{1}{\Delta_R})^{3-1}\nn\\
&=&\frac{i}{(4\pi)^2}\frac{1}{\Gamma[4]}\frac{1}{\Delta_R^2}
\label{identity1}
\eea

Identity 2:
\bea
&&\int \frac{d^4k}{(2\pi)^4}\frac{k_{||}^2}{[k^2_{||}-\Delta_{||}]^4} = 
i\,\int \frac{d^2k_{\perp}}{(2\pi)^2}\int \frac{d^2k_{E_{||}}}{(2\pi)^2}
\frac{-k^2_{E_{||}}}{[k^2_{E_{||}}+\Delta_{||}]^4}\nn\\
&=& -i\,\int \frac{d^2k_{\perp}}{(2\pi)^2}\frac{1}{4\pi}\frac{\Gamma[4-2]}{\Gamma[4]}(\frac{1}{\Delta_{||}})^2
= -i\, \frac{1}{(4\pi)^2}\frac{1}{\Gamma[4]}(\frac{1}{\Delta_R})\frac{\Gamma[1]}{\Gamma[2]}\nn\\
&=& -i\,\frac{1}{(4\pi)^2}\frac{1}{\Gamma[4]}\frac{1}{\Delta_R}
\label{identity2}
\eea
Therefore, using the above identities, we obtain first correction terms to the pion self- energy as mentioned already in Eq.(\ref{pi11}),
\bea
\Pi^{(1,1)}_{\pi^0}(q) = -\frac{{\rm g}^2_{\pi}}{4\pi^2}\int_0^1 dx\,x(1-x)
\Big[\frac{1}{\Delta_R}+\frac{m_p^2+x(1-x)q^2_{||}}{\Delta_R^2}\Big]
\eea
Next, we evaluate the term $T_6$. We start with the trace as follows,
\bea
T_6 &=& {\rm Tr}[\gamma_5S^{(2)}_a(k)\gamma_5S_b^{(0)}(k+q)]\nn\\
&=& {\rm Tr}[\gamma_5(k\!\!\!/ + m_{p}-\frac{\gamma.k_{\bot}}{k^2_{\bot}}(k^2-m^2_{p}))
\gamma_5(k\!\!\!/+q\!\!\!/ + m^2_{p})]\frac{-2k^2_{\bot}}{[k^2-m^2_{p}]^4[(k+q)^2-m^2_{p}]}\nn\\
&=& 4\Bigg[\big[k.(k+q)-m^2_p\big]
\frac{2k^2_{\bot}}{[k^2-m^2_p]^4[(k+q)^2-m^2_p]}+\big[k^2_{\bot}+k_{\bot}.q_{\bot}\big]
\frac{1}{[k^2-m^2_p]^3[(k+q)^2-m^2_p]}\Bigg]
\eea
Now, using the usual procedure of FP, we obtain,
\bea
T_6 &=& 4\Bigg[\f{[k^2+k.q(1-2x)+q^2x(x-1)-m_p^2][k_{\perp}^2-2xk_{\perp}.q_{\perp}+x^2q_{\perp}^2]}{[k_{||}^2-\Delta_{||}^2]^5}\nn\\
&+&\f{k_{\perp}^2+q_{\perp}^2x(x-1)+(k_{\perp}.q_{\perp})(1-2x)}{[k_{||}^2-\Delta_{||}^2]^4}\Bigg]
\eea
Similar to the evaluation of the $T_4$ term, we will make use of the following identities.

Identity 3:
\bea
\int \frac{d^4k}{(2\pi)^2} \frac{k^2k^2_{\perp}}{[k^2_{||}-\Delta_{||}]^5}=\frac{i}{(4\pi)^2}\frac{1}{\Gamma[5]}\frac{3}{\Delta_R}
\eea
Identity 4:
\bea
\int \frac{d^4k}{(2\pi)^4} \frac{k^2_{\perp}}{[k^2_{||}-\Delta_{||}]^5} &=& 
-i\,\int \frac{d^2k_{\perp}}{(2\pi)^2}k^2_{\perp}\frac{1}{4\pi}\frac{\Gamma[4]}{\Gamma[5]}\frac{1}{\Delta_{||}^4}\nn\\
&=& -\frac{i}{(4\pi)^2}\frac{\Gamma[4]}{\Gamma[5]}\frac{1}{\Gamma[4]}\frac{1}{\Delta_R^2}=
-\frac{i}{(4\pi)^2}\frac{1}{\Gamma[5]}\frac{1}{\Delta_R^2}
\eea
Identity 5:
\bea
\int \frac{d^4k}{(2\pi)^4} \frac{(k_{\perp}\cdot q_{\perp})^2}{[k^2_{||}-\Delta_{||}]^5} &=& 
-i\,\int \frac{d^2k_{\perp}}{(2\pi)^2}k^2_{\perp}\frac{q^2_{\perp}}{2}\frac{1}{4\pi}
\frac{\Gamma[4]}{\Gamma[5]}\frac{1}{\Delta_{||}^4}\nn\\
&=&-\frac{i}{(4\pi)^2}\frac{q^2_{\perp}}{2}\frac{1}{\Gamma[5]}\frac{1}{\Delta_R^2}
\eea
Identity 6:
\bea
\int \frac{d^4k}{(2\pi)^4} \frac{1}{[k^2_{||}-\Delta_{||}]^5} &=& 
-i\,\int \frac{d^2k_{\perp}}{(2\pi)^2}\frac{1}{4\pi}\frac{\Gamma[4]}{\Gamma[5]}\frac{1}{\Delta_{||}^4}\nn\\
&=&-\frac{i}{(4\pi)^2}\frac{\Gamma[4]}{\Gamma[5]}\frac{\Gamma[3]}{\Gamma[4]}\frac{1}{\Delta_R^3}
=-\frac{i}{(4\pi)^2}\frac{\Gamma[3]}{\Gamma[5]}\frac{1}{\Delta_R^3}
\eea
Identity 7:
\bea
\int \frac{d^4k}{(2\pi)^4} \frac{k_{\perp}^2}{[k^2_{||}-\Delta_{||}]^4} &=&
\frac{i}{(4\pi)^2}\frac{1}{\Gamma[4]}\frac{1}{\Delta_R}
\eea
Therefore, we obtain the second order correction to the pion self- energy as,
\bea
\Pi^{(2,0)}_{\pi^0}(q)
&=& -\frac{{\rm g}^2_{\pi}}{4\pi^2}
\Bigg[\int_0^1 dx(1-x)^3\big[\frac{1}{\Delta_R}+
\frac{q^2 x(1-x)+q^2_{\bot}x(4x-1)+m_p^2}{3\Delta_R^2}
+\frac{2x^2q_{\bot}^2[q^2x(1-x)+m_p^2]}{3\Delta_R^3}\big]\nn\\ 
&&~~~~~~+\int_0^1 dx(1-x)^2\big[\frac{1}{\Delta_R}
-\frac{q^2_{\bot}x(1-x)}{\Delta_R^2}\big]\Bigg]\label{pi200}.
\eea
\section*{Appendix C: Details of calculation of self energy for pseudo- vector coupling}
The same formalism for calculation of the correction terms in PS coupling can be applied to the PV coupling. So, we start by calculation of the traces of the terms in the self energy as,
\bea
&&{\rm Tr}\Big[\gamma_5q\!\!\!/S_a(k)\gamma_5q\!\!\!/S_b(k+q)\Big]\nn\\
&=&{\rm Tr}\Bigg[\gamma_5q\!\!\!/S^{(0)}_a(k)\gamma_5q\!\!\!/S_b^{(0)}(k+q)
+(e B)\Big\{\gamma_5q\!\!\!/S^{(0)}_a(k)\gamma_5q\!\!\!/S_b^{(1)}(k+q)
+\gamma_5q\!\!\!/S^{(1)}_a(k)\gamma_5q\!\!\!/S_b^{(0)}(k+q)\Big\}\nn\\
&+&(e B)^2\Big\{\gamma_5q\!\!\!/S^{(1)}_a(k)\gamma_5q\!\!\!/S_b^{(1)}(k+q)+
\gamma_5q\!\!\!/S^{(0)}_a(k)\gamma_5q\!\!\!/S_b^{(2)}(k+q)+
\gamma_5q\!\!\!/S^{(2)}_a(k)\gamma_5q\!\!\!/S_b^{(0)}(k+q)\Big\}+\mathcal{O}((e B)^3)\Bigg]\nn\\
&\simeq& T_1' + (e B)[T_2' + T_3'] + (e B)^2[T_4' + T_5' + T_6']
\eea  
where,
\bea
T_1' &=& {\rm Tr}[\gamma_5q\!\!\!/S^{(0)}_a(k)\gamma_5q\!\!\!/S_b^{(0)}(k+q)]\nn\\
T_2' &=& {\rm Tr}[\gamma_5q\!\!\!/S^{(0)}_a(k)\gamma_5q\!\!\!/S_b^{(1)}(k+q)]\nn\\
T_3' &=& {\rm Tr}[\gamma_5q\!\!\!/S^{(1)}_a(k)\gamma_5q\!\!\!/S_b^{(0)}(k+q)]\nn\\
T_4' &=& {\rm Tr}[\gamma_5q\!\!\!/S^{(1)}_a(k)\gamma_5q\!\!\!/S_b^{(1)}(k+q)]\nn\\
T_5' &=& {\rm Tr}[\gamma_5q\!\!\!/S^{(0)}_a(k)\gamma_5q\!\!\!/S_b^{(2)}(k+q)]\nn\\
T_6' &=& {\rm Tr}[\gamma_5q\!\!\!/S^{(2)}_a(k)\gamma_5q\!\!\!/S_b^{(0)}(k+q)]
\eea

 $T_4'$ is evaluated as,
\bea
T_4' &=& {\rm Tr}\Bigg[\gamma_5q\!\!\!/S_p^{(1)}(k)\gamma_5q\!\!\!/S_p^{(1)}(k+q)\Bigg] \nn\\
&=&{\rm Tr} \Bigg[\gamma_5q\!\!\!/i\gamma_1\gamma_2(\gamma.k_{||}+m_p)
\gamma_5q\!\!\!/i\gamma_1\gamma_2(\gamma.(k+q)_{||}+m_p)\Bigg]
\frac{1}{[k^2-m_p^2]^2[(k+q)^2-m_p^2]^2}\nn\\
&=&\f{4}{\Delta_4}\Bigg[q^2(k_{||}^2+(k_{||}\cdot q_{||})) + 2 (k_0 q_3-k_3 q_0)^2+q^2 -2 q^2_{||}\Bigg]
\eea
where $\Delta_4=[k^2-m_p^2]^2[(k+q)^2-m_p^2]^2$. Using FP (and substituting $k\rightarrow (k-xq)$), we obtain,
\bea
T_4'=\f{4}{[k^2_{||}-\Delta_{||}]^4}\Bigg[q^2[k_{||}^2+x(x-1)q_{||}^2+(1-2x)(k_{||}\cdot q_{||})]+2 (k_0 q_3-k_3 q_0)^2+q^2 -2 q^2_{||}\Bigg]
\eea
where, $\Delta_{||} = k^2_{\perp}+\Delta_R$.
Therefore, we need to evaluate the integral,
\bea
&&\int \frac{d^4k}{(2\pi)^4} T_4 \nn\\
&=&4\int_0^1x(1-x)\,dx\Gamma[4]\int \frac{d^4k}{(2\pi)^4}
\frac{q^2[k_{||}^2+x(x-1)q_{||}^2+(1-2x)(k_{||}\cdot q_{||})]+2 (k_0 q_3-k_3 q_0)^2
+m_p^2(q^2 -2q^2_{||})}{[k^2_{||}-\Delta_{||}]^4}
\eea

Identity 8:
\bea
&&\int^{\infty}_{-\infty} \int^{\infty}_{-\infty}\frac{dk_0\,dk_3}{(2\pi)^2}\frac{k^2_0}{(k^2_{||}-\Delta_{||})^4}
=i\,\frac{1}{4\pi}\frac{1}{\Gamma[4]}\frac{1}{2\Delta_{||}^2}\\
\eea
Identity 9:
\bea
&&\int^{\infty}_{-\infty} \int^{\infty}_{-\infty}\frac{dk_0\,dk_3}{(2\pi)^2}\frac{k^2_3}{(k^2_{||}-\Delta_{||})^4}
=-i\,\frac{1}{4\pi}\frac{1}{\Gamma[4]}\frac{1}{2\Delta_{||}^2}\\
\eea
Identity 10:
\bea
&&\int^{\infty}_{-\infty} \int^{\infty}_{-\infty}\frac{dk_0\,dk_3}{(2\pi)^2}\frac{k_0k_3}{(k^2_{||}-\Delta_{||})^4} = 0\\
\eea
Identity 11:
\bea
&&\int \frac{d^2k_{\perp}}{(2\pi)^2}\frac{1}{(k^2_{\perp}+\Delta_R)^2} = \frac{1}{4\pi}\frac{1}{\Delta_R} 
\eea
Therefore, using the above identities, we obtain first correction terms to the pion self- energy as,
\bea
\Pi^{(1,1)}_{\pi^0}(q) = (\frac{f_{\pi}}{m_{\pi}})^2\frac{1}{4\pi^2}\int_0^1x\,(1-x)\,dx\,
\Big[\frac{x(x-1)q^2(2q_{||}^2-q^2)-m^2q^2_{||}}{\Delta_R^2}\Big].
\eea
As we have not differentiated between the masses of neutron and proton, the terms $T_5'$ and $T_6'$ are identical.
We start by evaluating the trace in the $T_6'$ term,
\bea
{\rm T}_6' &=& {\rm Tr}\Big[\gamma_5q\!\!\!/S_p^{(2)}(k)\gamma_5q\!\!\!/S_p^{(0)}(k+q)\Big]\nn\\
&=& {\rm Tr}\Big[\gamma_5q\!\!\!/[(k\!\!\!/+m_p)-
\frac{\gamma\cdot k_{\perp}}{k_{\perp}^2}(k^2-m_p^2)]\gamma_5 q\!\!\!/[(k\!\!\!/+q\!\!\!/)+m_p]\Big]
\frac{-2k^2_{\perp}}{(k^2-m_p^2)^4((k+q)^2-m_p^2)}\nn\\
&=&\frac{-2k^2_{\perp}}{\Delta_6}{\rm Tr}\Bigg[\{\gamma_5q\!\!\!/k\!\!\!/\gamma_5q\!\!\!/(k\!\!\!/+q\!\!\!/)
+\gamma_5q\!\!\!/\gamma_5q\!\!\!/ m_p^2\}-
\{\gamma_5q\!\!\!/(\gamma\cdot k_{\perp})\gamma_5q\!\!\!/(k\!\!\!/+q\!\!\!/)\frac{k^2-m_p^2}{k^2_{\perp}}\}\Bigg]
\label{seccorr1}
\eea
where $\Delta_6=(k^2-m_p^2)^4((k+q)^2-m_p^2)$. The first term in $\{\}$ in Eq.(\ref{seccorr1}) gives,
\bea
&&{\rm Tr}\Big[\gamma_5q\!\!\!/k\!\!\!/\gamma_5q\!\!\!/(k\!\!\!/+q\!\!\!/)
+\gamma_5q\!\!\!/\gamma_5q\!\!\!/ m_p^2\Big]k^2_{\perp} = 4\Big [2(k.q)^2+q^2(k.q)-q^2k^2-q^2m_p^2\Big]k^2_{\perp}\nn\\
&=&4\Bigg[2(k\cdot q)^2 k^2_{\perp}-q^2k^2k^2_{\perp}+2x^2(k\cdot q)^2q^2_{\perp}-x^2q^2k^2q^2_{\perp}+
q^2[x(x-1)q^2-m_p^2]k^2_{\perp}\nn\\
&-&2x(1-2x)q^2(k\cdot q)(k_{\perp}\cdot q_{\perp})+x^2q^2q^2_{\perp}[x(x-1)q^2-m_p^2]+(1-2x)q^2(k\cdot q)k^2_{\perp}\nn\\
&+&(1-2x)x^2q^2(k\cdot q)q^2_{\perp}-4x(k\cdot q)^2(k_{\perp}\cdot q_{\perp})+2xq^2k^2(k_{\perp}\cdot q_{\perp})
-2xq^2[x(x-1)q^2-m_p^2](k_{\perp}\cdot q_{\perp})
\eea
where we have used FP by using $k\rightarrow (k-xq)$.
In a similar way, following the procedure of FP, the second term in $\{\}$ in Eq.(\ref{seccorr1}) gives,
\bea
&&{\rm Tr}\Big[\gamma_5q\!\!\!/(\gamma\cdot k_{\perp})\gamma_5q\!\!\!/(k\!\!\!/+q\!\!\!/)\Big] \nn\\
&=&4\Bigg[2(q\cdot k)(q\cdot k_{\perp})-q^2(k\cdot k_{\perp}) + q^2(q\cdot k_{\perp})\Bigg]\nn\\
&=&4\Bigg[2(k_{\perp}\cdot q_{\perp})^2+q^2k_{\perp}^2-x(x-1)q^2q_{\perp}^2-2(k_{||}\cdot q_{||})(k_{\perp}\cdot q_{\perp})
+2x(k\cdot q)q_{\perp}^2-q^2(k_{\perp}\cdot q_{\perp})\Bigg]
\eea
Now, the denominator in Eq.(\ref{seccorr1}) is also modified by the momentum substitution as,
\bea
\frac{1}{\Delta_6} &=& \frac{1}{(k^2-m_p^2)^4((k+q)^2-m_p^2)} \nn\\
&=&\frac{1}{[k^2_{||}-\Delta_{||}]^5}
\eea
We have used the following identities to arrive at the second correction term. They are listed as follows,

Identity 12:
\bea
\int \frac{d^4k}{(2\pi)^4} \frac{(k\cdot q)^2k^2_{\perp}}{[k^2_{||}-\Delta_{||}]^5} &=& 
\frac{q^2}{4}\int \frac{d^4k}{(2\pi)^4} \frac{k^2k^2_{\perp}}{[k^2_{||}-\Delta_{||}]^5}=
\frac{q^2}{4}\int \frac{d^4k}{(2\pi)^4}\frac{(k^2_{||}-k^2_{\perp})k^2_{\perp}}{[k^2_{||}-\Delta_{||}]^5}\nn\\
&=& \frac{i}{(4\pi)^2}\frac{q^2}{4}\frac{1}{\Gamma[5]}\frac{3}{\Delta_R}
\eea
Identity 13:
\bea
\int \frac{d^4k}{(2\pi)^2} \frac{k^2k^2_{\perp}}{[k^2_{||}-\Delta_{||}]^5}=\frac{i}{(4\pi)^2}\frac{1}{\Gamma[5]}\frac{3}{\Delta_R}
\eea
Identity 14:
\bea
\int \frac{d^4k}{(2\pi)^4} \frac{(k\cdot q)^2}{[k^2_{||}-\Delta_{||}]^5}&=& 
\frac{q^2}{d}\int \frac{d^4k}{(2\pi)^4} \frac{k^2}{[k^2-\Delta_R]^5}\nn\\
&=&\frac{i}{(4\pi)^2}\frac{q^2}{4}\frac{\Gamma[3]}{\Gamma[5]}\frac{1}{\Delta_R^2}
\eea
Identity 15:
\bea
\int \frac{d^4k}{(2\pi)^4} \frac{k^2}{[k^2_{||}-\Delta_{||}]^5} = \frac{i}{(4\pi)^2}\frac{\Gamma[3]}{\Gamma[5]}\frac{1}{\Delta_R^2}
\eea
Identity 16:
\bea
\int \frac{d^4k}{(2\pi)^4} \frac{(k_{\perp}\cdot q_{\perp})^2}{[k^2_{||}-\Delta_{||}]^4} &=& 
i\int \frac{d^2k_{\perp}}{(2\pi)^2}k^2_{\perp}\frac{q^2_{\perp}}{2}
\frac{1}{4\pi}\frac{\Gamma[3]}{\Gamma[4]}\frac{1}{\Delta^3_{||}}\nn\\
&=& \frac{i}{(4\pi)^2}\frac{q^2_{\perp}}{2}\frac{\Gamma[3]}{\Gamma[4]}\frac{1}{\Gamma[3]}\frac{1}{\Delta_R} = 
\frac{i}{(4\pi)^2}\frac{q^2_{\perp}}{2}\frac{1}{\Gamma[4]}\frac{1}{\Delta_R}
\eea
Usin the above identities in addition to Identities [4-7], we obtain the second correction term to the pion self energy as,
\bea
\Pi^{(2,0)}_{\pi^0}(q) &=& (\frac{f_{\pi}}{m_{\pi}})^2\frac{1}{4\pi^2}\Bigg[\int_0^1dx\frac{(1-x)^3}{3}\,q^2\,\Big[\frac{3}{2\Delta_R}-
x(x-1)\frac{q^2_{\perp}}{\Delta_R^2}+[x(x-1)q^2-m^2_p](\frac{1}{\Delta_R^2}+2x^2\frac{q^2_{\perp}}{\Delta_R^3})\Big]\nn\\
&+&\int_0^1dx(1-x)^2\Big[\frac{q^2+q^2_{\perp}}{\Delta_R}-x(x-1)\frac{q^2q^2_{\perp}}{\Delta_R^2}\Big]\Bigg] 
\eea
\subsection*{Acknowledgements}
S.P.A. would like to thank Prof. Alejandro Ayala for valuable discussions and suggestions regarding different aspects of this work.

\end{document}